\documentclass[aps,prb,twocolumn,showpacs,floatfix,superscriptaddress]{revtex4-1}

\usepackage{graphicx,color}
\usepackage{amsfonts}
\usepackage[figuresright]{rotating}  
\usepackage{amssymb}
\usepackage{amsmath}
\usepackage{mathtools}
\usepackage{psfrag}
\usepackage{subfigure}
\usepackage{multirow}
\usepackage{tabularx}
\usepackage{textcomp}
\usepackage{bm}
\usepackage{hyperref}
\usepackage{relsize}
\hypersetup{
 pdfnewwindow=true, colorlinks=true,
 linkcolor=blue, anchorcolor=blue,
 citecolor=blue, filecolor=blue,
 menucolor=blue, urlcolor=blue}

\def\beq{\begin{eqnarray}}
\def\eeq{\end{eqnarray}}

\renewcommand{\v}[1]{\ensuremath{\mathbf{#1}}} 
 
\let\baraccent=\= 
\renewcommand{\=}[1]{\stackrel{#1}{=}} 

\makeatletter

\begin{document}

\title{Quantitative relationship between polarization differences and the zone-averaged shift photocurrent}

\author{Benjamin\ \surname{M. Fregoso}}
\affiliation{Department of Physics, University of California, Berkeley,
California, 94720, USA}

\author{Takahiro\ \surname{Morimoto}}
\affiliation{Department of Physics, University of California, Berkeley,
California, 94720, USA}

\author{Joel\ \surname{E. Moore}}
\affiliation{Department of Physics, University of California, Berkeley,
California, 94720, USA}
\affiliation{Materials Sciences Division, Lawrence Berkeley National Laboratory, Berkeley, CA 94720}

\begin{abstract}
A relationship is derived between differences in electric polarization between bands and the ``shift vector'' 
that controls part of a material's bulk photocurrent, then demonstrated in several models.  Electric 
polarization has a quantized gauge ambiguity and is normally observed at surfaces via the surface charge density, 
while shift current is a bulk property and is described by shift vector gauge invariant at each point in momentum 
space.  They are connected because the same optical transitions that are described in shift currents pick out a 
relative gauge between valence and conduction bands.  We also discuss subtleties arising when there are 
points at the Brillouin zone where optical transitions are absent. We conclude that two-dimensional materials with 
significant interband polarization differences should have high bulk photocurrent, meaning that the modern 
theory of polarization can be used as a straightforward way to search for bulk photovoltaic material candidates.
\end{abstract}

\maketitle

\section{Introduction} 
Many electronic and optical properties of crystals depend not just on 
the energy band structure but on the detailed properties of Bloch 
wave functions.  A simple example is that optical transitions in a solid, 
just like in an atom, involve matrix elements that depend on the 
symmetries of the underlying wave functions or orbitals.  A deeper example 
is that the geometric or Berry phase of Bloch wave functions controls the 
electrical polarization and other properties.  Although the spontaneous 
polarization of solids was already of interest to the ancients, and the 
polarization of a finite distribution of charge density is easily 
understood, the proper computation of electrical polarization from a unit 
cell of an infinite crystal had to await the ``modern theory of 
polarization'',~\cite{Thouless83,King-Smith1993,VK93,Resta1994} which is now widely used in 
practical calculations.

The goal of the present paper is to explain the quantitative connection 
between bulk nonlinear optical properties of a material, specifically the 
shift current piece of photocurrent linear in the intensity of applied 
light, and electrical polarization. The shift current response 
is determined by a third rank tensor, 
\begin{align}
J^a_{shift}= 2\sum_{b}\sigma^{abb} E^{b}(\omega) E^{b}(-\omega),
\end{align}
where the electric field is $E^{b}(t)=E^{b}(\omega)e^{-i\omega t} + E^{b}(-\omega)e^{i\omega t}$.
It is nonvanishing when inversion symmetry is absent, e.g., for ferroelectric materials. The tensor 
can be written in an intuitive way as (see Appendix~\ref{app:phase_ind_shiftv})
\begin{align}
\sigma^{abb} \approx \frac{e}{\hbar }\sum_{nm}\int_{\textit{{\tiny BZ}}} R_{nm}^{a,b} \varepsilon^{bb}_{2,nm} ,
\label{supp:shift_tensor_main_text}
\end{align}
where $\varepsilon^{bb}_{2,nm}(\v{k},\omega)$ is the diagonal (band-resolved) 
imaginary part of the dielectric function, which is proportional to the density 
of states, and $\int_{\textit{{\tiny BZ}}} \equiv \int d\v{k}/(2\pi)^d$ 
represents an integral over the Brillouin zone  (BZ) in $d$ dimensions. 
In the following we often suppress the frequency and momentum dependence of 
quantities for simplicity of notation. Importantly, the shift current includes a geometrical 
{\it shift  vector} $R^{a,b}_{nm}$~\cite{Sturman1992,Baltz1981,Sipe2000,YR12,Young-Zheng-Rappe}
defined by, 
\begin{align}
R^{a,b}_{nm} =\frac{\partial \phi^{b}_{nm}}{\partial k^a} + A^a_{nn} -A^a_{mm},
\label{eqn:shift_vec_multiband}
\end{align}
where $A_{nm}^b$ are the Berry connections
\begin{align}
A_{nm}^b = i \langle u_n |  \frac{\partial}{\partial k^b} | u_m \rangle,
\end{align}
and $u_n$ is the periodic part of the Bloch wave function at wave vector $\v{k}$.
$b=x,y,z$ is a Cartesian axis, and $\phi_{nm}^b$ is the phase of the connection 
$A^b_{nm}=|A^b_{nm}|e^{-i \phi_{nm}^b}$. The shift vector also determines the second harmonic 
generation and electro-optic responses~\cite{Sipe2000,Morimoto16} of 
semiconductors.
 
We note that the definition of shift vector in (\ref{eqn:shift_vec_multiband}) involves the gauge-dependent 
quantities $A^b_{nn}, A^b_{mm}$, and $\phi^b_{nm}$.  However, the combination is 
gauge invariant, at all points of the BZ where the optical transition matrix element $A^b_{nm}$ is nonzero. 
Conversely, electrical polarization is written in the standard 
theory as an integral of the locally gauge-dependent Berry connection.
In other words, the contribution of a particular $k$-point to the electrical 
polarization is not meaningfully defined.  The total polarization is 
gauge dependent up to a quantized ambiguity; in the simplest case of one 
spatial dimension, the polarization
\begin{align}
e\int_{\textit{{\tiny BZ}}}A_{nn} = P_n 
\end{align}
is defined only up to addition of an integer multiple of electron charge. For example, gauge 
transformations $u_n \rightarrow e^{i \varphi_n} u_n$ change $P_n$ by $j e$, where $j \in \mathbb{Z}$ 
is the winding number of the angular variable $\varphi_n$ around the BZ. The physical bulk polarization 
is defined as a difference with respect to an inversion-symmetric reference system which is adiabatically 
deformed with each other while keeping a fixed value of $j$. Nevertheless, the (gauge-invariant) shift 
vector is directly related in many cases to (gauge-dependent) polarization differences between the valence 
and conduction band.   

The shift current mechanism has recently gained interest for its potential novel optoelectronic applications based on 
ferroelectrics~\cite{Tan2016a,Wang2016,Grinberg,Nie,Shi,deQuilettes}. In particular, 2D materials have highly tunable 
electronic and optical properties~\cite{Gomes2016,Mehboudi2016,Naumis2016,Salazar2016} and are expected to 
generate large shift current~\cite{Zenkevich2014,cook_design_2015,Rangel2017}. We can identify 
three factors that determine the magnitude of the shift current: density of states, velocity matrix elements, and
shift-vector matrix elements. In three dimensions, they are all intertwined with no obvious relation among
them~\cite{YR12}. In two dimensions, on the other hand, the density of states is constant and the optical transitions are 
determined by velocity and shift-vector matrix elements. Approximating the dipole matrix 
elements $|r_{nm}^b|^2$ by a constant $\varepsilon^{bb}$ [see Appendix~\ref{app:phase_ind_shiftv}
Eq.~(\ref{supp:shift_tensor_gen})] we obtain
\begin{align}
\sigma^{abb}\approx -\frac{\pi e^3 \varepsilon^{bb}}{\hbar^2}\sum_{nm}\int_{\textit{{\tiny BZ}}} f_{nm}
R_{nm}^{a,b} \delta(\omega_{mn}-\omega),
\label{supp:shift_tensor_main_text_v2}
\end{align}
where $\hbar\omega_{nm}=\hbar\omega_{n}-\hbar\omega_{m}$  are band energy differences
and $f_{nm}=f_n - f_m$ differences of Fermi distribution functions of band $n$ and 
$m$. As pointed out in Ref.~\onlinecite{Rangel2017}, in real-life applications such as solar cells,
the integrated response over a frequency range is more important than the 
response at a single frequency. Integrating over all frequency,
\begin{align}
\int d\omega~ \sigma^{abb}\approx -\frac{\pi e^3 \varepsilon^{bb}}{\hbar^2}\sum_{nm}\int_{\textit{{\tiny BZ}}} f_{nm}
R_{nm}^{a,b},
\label{supp:shift_tensor_main_text_v3}
\end{align}
we see that the total short-circuit current is proportional to the integrated shift vector over the BZ. 
As shown below, the integral of the shift vector over the BZ is equal to the polarization difference evaluated 
in a specific gauge; the optical transitions mediated by the shift vector can be viewed as fixing the relative 
gauge between valence and conduction bands, at least in the simplest case where such transitions are 
allowed at every $k$-point. This connection between polarization and shift vector indicates that materials with 
significant polarization differences between bands (minimized over gauge ambiguities) must have significant shift 
vectors somewhere in the BZ. In order to understand this relation we consider simple models first.

\begin{figure}[]
\subfigure{\includegraphics[width=0.48\textwidth]{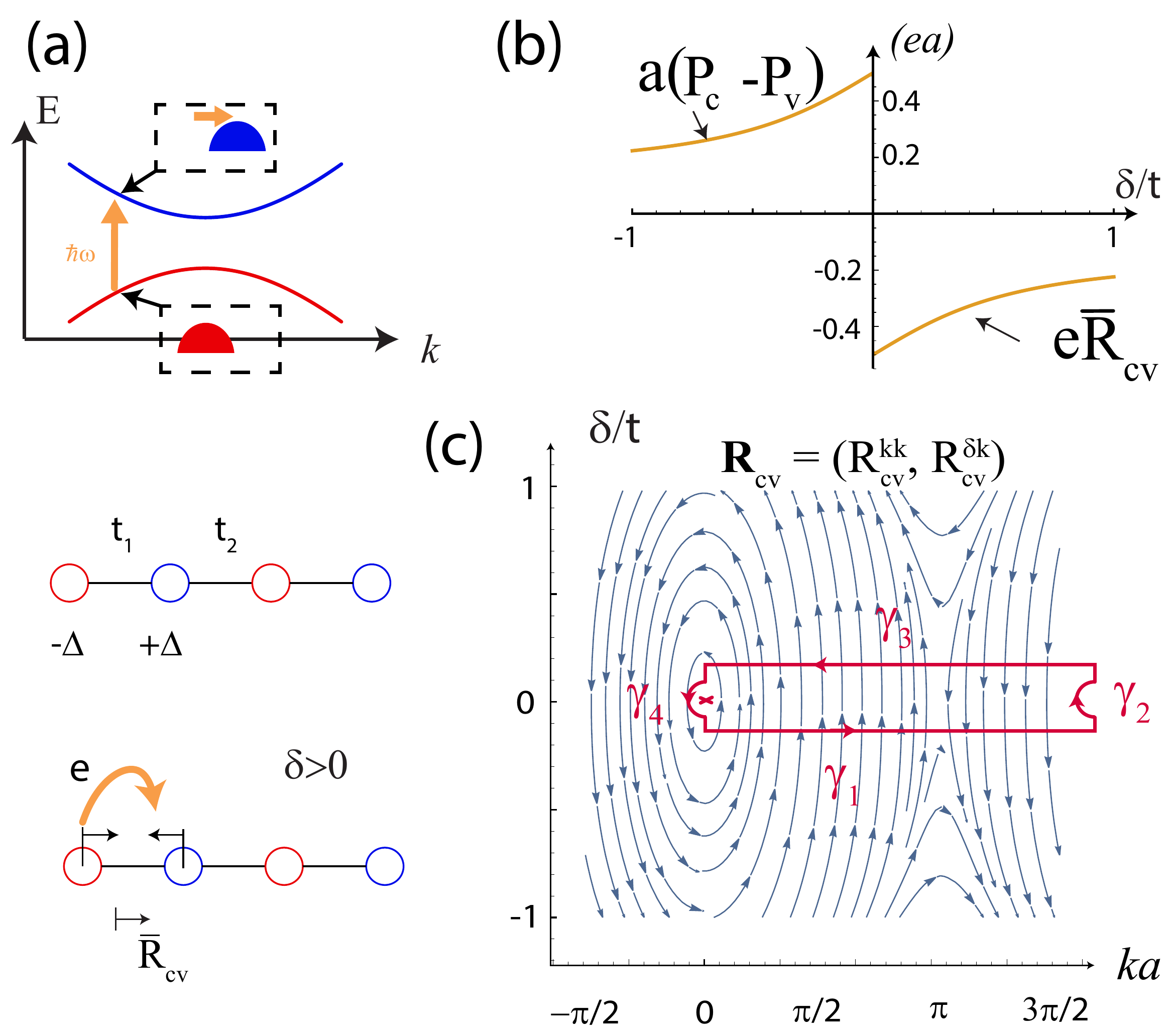}}
\caption{(a) Top panel: Photoexcitation induces shift of the electron wavepacket in real space. 
(a) Bottom panel: Rice-Mele (RM) tight-binding model. The unit cell of size $a$ has two sites and
alternating hoppings $t_1=t/2+\delta/2$ and $t_2=t/2-\delta/2$.
The distance between conduction and valence band centers is $\bar{R}_{cv}$. 
For $\delta=0$, $\bar{R}_{cv}=\pm a/2$ is ambiguous because the system does not break inversion symmetry.
For $\delta>0$, the centers of charge move towards one another by a distance $d$. 
The polarization is $P_v(\delta)- P_v(0) = -ed = (\bar{R}_{cv}-a/2)/2$.
When a photon is absorbed the electron jumps to another atom a distance $\bar{R}_{cv}$ 
away.
(b) Integral of the shift vector over the BZ and polarization difference. 
$\bar{R}_{cv}$ has an integer discontinuity at $\delta=0$.
(c) Stream plot of the vector field $\v{R}_{cv}=(R_{cv}^{kk},R_{cv}^{k\delta})$ which has vortex 
of charge $+1$ in this \textit{gauge-independent} vector field (see main text). 
The discontinuity in $\bar{R}_{cv}$ is the charge of the optical zero.
In the numerical examples $\Delta>0$ and $t=e=a=1$.}
\label{fig:shift_vector_RM_model}
\end{figure}

\section{Relationship between shift vector and polarization}
We start our analysis focusing on one-dimensional (1D) systems. Let us consider conduction and valence 
bands, which we label with $c$ and $v$, separated by an energy gap. In particular, we consider 
insulators with broken inversion symmetry that support nonzero polarization, where the wave functions 
and off-diagonal Berry connections are complex. In addition, we adopt the periodic gauge~\cite{Martin2008,Lax1974} 
defined by $\psi_{n}(k+G,r)=\psi_{n}(k,r)$ where $\psi_n$ are Bloch wavefunctions and $G$ a reciprocal 
lattice vector. In this case, all connections $A_{nm}$ are periodic in the BZ, i.e., $A_{cv}(k+G)=A_{cv}(k)$ 
(see Appendix~\ref{app:optical_gauge}). Since the phases $\phi_{cv}$ at $k$ and $k+G$ coincide modulo $2\pi$, we can 
define winding $W_{cv}$ of the phase $\phi_{c,v}$ around the BZ as
\begin{align}
W_{cv} &= \frac{1}{2\pi} \oint d\phi_{cv} \in \mathbb{Z}.
\label{eq:Wcv_def}
\end{align}
Here the winding $W_{cv}$ can be any integer because we still have the freedom to perform 
transformations such that $\partial_k \varphi_n$ is periodic, e.g., large gauge transformations 
that change the value of $W_{cv}$ and keep $\psi_{n}$ periodic over the BZ.
We define the \textit{optical gauge} by further constraining the periodic gauge 
such that $\phi_{cv}=0$ and constant. When $A_{cv}=0$ at some $k$-point in the BZ 
(which we call ``optical zero''), the phase $\phi_{cv}$ is not well-defined, and hence, 
$W_{cv}$ is multivalued. The existence of optical zeros is physical and 
cannot be removed by gauge transformations.

Since $R_{cv}$ is related to the shift of wave packets [see Fig.~\ref{fig:shift_vector_RM_model}(a)], we can expect that 
an integral of $R_{cv}$ over $k$ has a relationship to the difference of polarization of 
the two bands. Indeed, integrating $R_{cv} =\partial_k \phi_{cv} + A_{cc} -A_{vv}$, leads to 
\begin{align}
e\bar{R}_{cv} &= ea \int_{\textit{{\tiny BZ}}} R_{cv} = e a W_{cv} +a P_c - a P_v,
\label{eqn:int_shift_vector}
\end{align}
where $P_{c,v}$ is polarization of conduction and valence bands. Eq.~(\ref{eqn:int_shift_vector}) 
shows that the integral of the shift vector over the BZ is proportional to the polarization   
difference between the conduction and valence bands up to an integer $W_{cv}$.
In particular, the optical gauge allows us to directly connect shift vector and polarization as
\begin{align}
e\bar{R}_{cv} = a P_c - a P_v
\end{align}
since $W_{cv}=0$. We emphasize that this is only possible when there is no optical zero  in the 
region of the integral. Let us consider some explicit examples.
\begin{figure}[]
\subfigure{\includegraphics[width=0.47\textwidth]{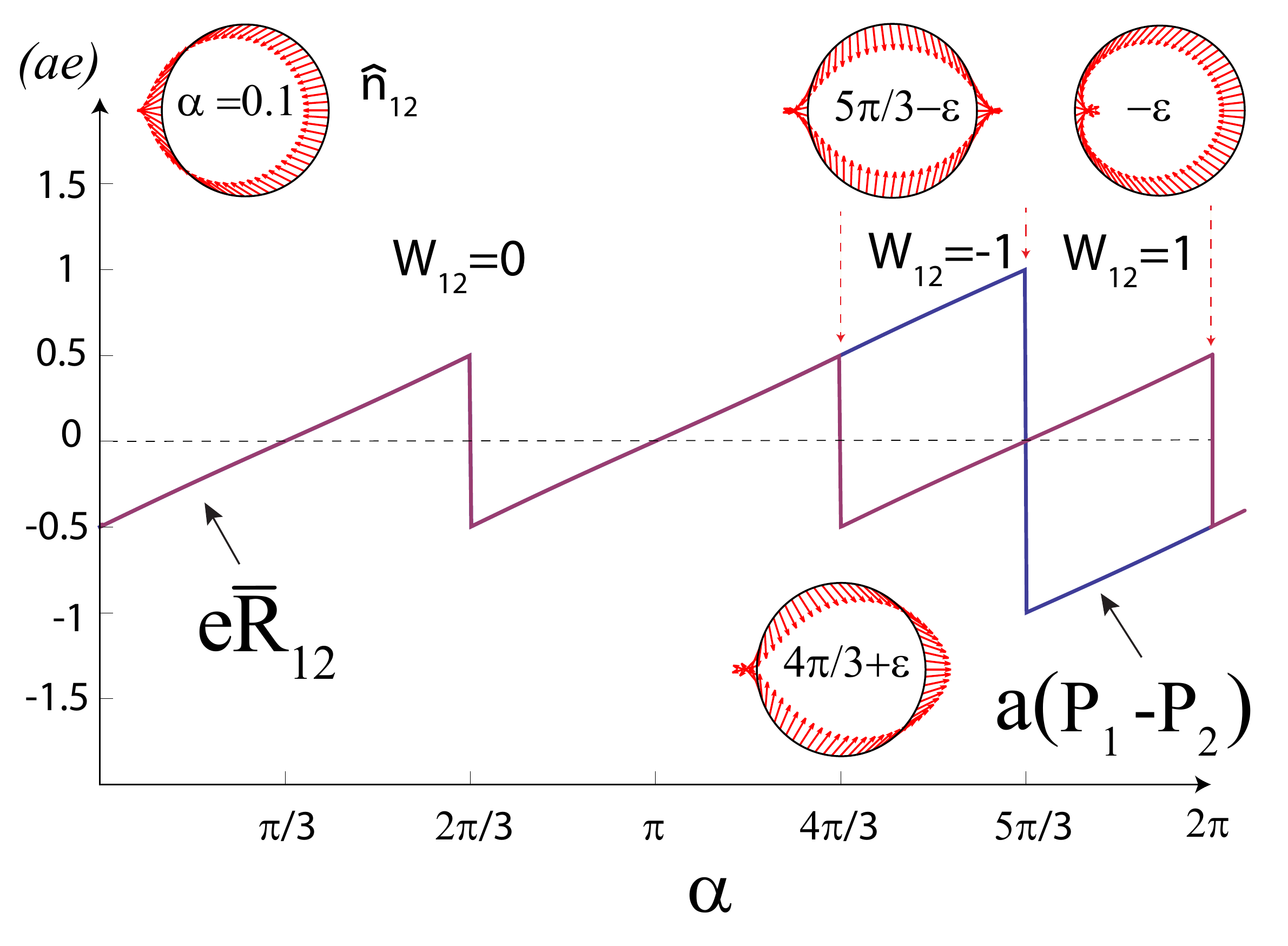}}
\caption{Polarization and integrated shift vector in a three-band model, Eq.~(\ref{eq:3B_model_ham}). 
We find jumps in $W_{12}$ indicating that no single gauge choice gives vanishing winding numbers over the 
parameter $\alpha$. However, $e \bar{R}_{12} = a(P_1 - P_2) + W_{12} e a$ holds for all $\alpha$.
We used parameters $B/A=0.5, e=a=1$ and $0<\varepsilon\ll 1$. In evaluating $P_1-P_2$, we adopted 
the gauge given in Eq.~(\ref{eq.wave_func}) with $\varphi_n=0$. }
\label{fig:Vector_field}
\end{figure}

\section{Rice-Mele model}
Let us apply the above analysis to the Rice-Mele (RM) model~\cite{RM82} [see Fig.~\ref{fig:shift_vector_RM_model}(a)], 
which is an archetypal model of ferroelectricity along the polar axis. It applies to polyacetylene, 
BaTiO$_3$, and even monochalcogenides\cite{Rangel2017}. It is given by
\begin{align}
\hat{H}= \sum_{i} [ (\frac{t}{2} + 
(-1)^{i}\frac{\delta}{2}) c_{i}^{\dagger}c_{i+1} + h.c. + (-1)^{i} \Delta c^{\dagger}_{i} c_{i} ].
\end{align}
The $c_i$ ($c^{\dagger}_i$) annihilates (creates) s-wave electron states at site $i$. The unit cell of 
size $a$ has two sites; $\delta$ parametrizes the dimerization of the chain and $\Delta$ the staggered 
on-site potential, Fig.~\ref{fig:shift_vector_RM_model}(a). Inversion symmetry is broken if 
$\Delta\neq 0$ and $\delta\neq 0$ and preserved otherwise 
(for details of the model, see Appendix~\ref{sec:shift_vector_RM}). The shift vector 
for this (and any two-band) model can be computed and studied analytically.
For example its gauge invariance is made apparent when we write it in terms of the Hamiltonian 
and its derivatives (see Appendix~\ref{app:two-band_model}).

With a gauge in Eq.~\ref{eq:wf_RM_model}, we obtain $W_{cv}=0$ [see Fig.~\ref{fig:shift_vector_RM_model}(b)]. 
The shift vector $R_{cv}$ is usually assigned the meaning of the size of the microscopic dipole formed 
by the photo-excited electron~\cite{Sturman1992}. Since $R_{cv}$ could grow without limit 
(see Appendix~\ref{sec:shift_vector_RM}), we believe, $\bar{R}_{cv}$ has a more well defined 
physical meaning, namely, as the distance between the valence and conduction 
centers of charge [Fig.~\ref{fig:shift_vector_RM_model}(a)], and is therefore bounded by the lattice spacing $a$. 

At $(ka,\delta)=(0,0)$ we have $A_{cv}=0$, and the size of the 
discontinuity in $\bar{R}_{cv}$  [Fig.~\ref{fig:shift_vector_RM_model}(b)] 
is determined by the vorticity associated with the optical zero as follows. 
We consider the parameter $\delta$ as if it were a Cartesian direction 
and define the gauge invariant shift vectors as
$R^{\mu\nu}_{cv} = \partial_{\mu} \phi_{cv}^{\nu} + A_{cc}^{\mu} - A_{vv}^{\mu}$ with 
$\mu,\nu=ka,\delta$. The singularity at the optical zero is clear in
the vector field $\v{R}_{cv}= (R_{cv}^{kk},R_{cv}^{\delta k})$ shown in 
Fig.~\ref{fig:shift_vector_RM_model}(c). At the optical zero, 
the shift vector diverges (for details, see Appendix~\ref{sec:shift_vector_RM}). The jump in 
$\bar{R}_{cv}^{kk} (=\bar{R}_{cv}$) at $\delta=0$ is obtained from the 
integral of $\v{R}_{cv}$ along the path $\gamma=\sum^4_{n=1}\gamma_n$, which leads to
\begin{align}
\frac{1}{2\pi}\oint_{\gamma_2\gamma_4 \to  0} \hspace{-15pt}d \v{ \lambda} \cdot  \v{R}_{cv} = \frac{\bar{R}_{cv}(0^{-}) -\bar{R}_{cv}(0^{+})}{a} =1,
\end{align}
with $d\v{\lambda} \equiv (dk, d\delta)$.
One can check that the vortex at $(k a,\delta)=(\pi,0)$ does not 
contribute to the path integral since $\v{R}_{cv}$ vanishes at 
this point. Furthermore, this vortex structure at optical zeros 
governs the charge pumping induced by a periodic change of parameter 
[e.g., over a path $(\Delta,\delta)=(\cos\theta,\sin\theta)$ with 
$\theta=0 \to 2\pi$]. The pumped charge in this circuit is given 
by ``the Berry curvature'' as $\int_{S} \Omega_{\mu\nu}^{cv}$ with 
$\Omega_{\mu\nu}^{cv}=\partial_{\mu} R_{cv}^{\mu k} - \partial_{\nu} R_{cv}^{\nu k}$.

\section{1D three-band model with inversion breaking}
Next we show that the direct relationship between shift vector and polarization is not limited 
to the two band models by demonstrating the relationship in the case of general number of bands. 
As an example, we consider the three-band model described by
\begin{align}
\hat{H}=\sum_j t_j c^{\dagger}_j c_{j+1} + h.c.,
\label{eq:3B_model_ham}
\end{align}
with $t_j= A+ B \cos{\big(2\pi j/3 -\alpha\big)}$. In this model the lower band pumps $-2e$ while 
the other two pump $e$ per cycle in $\alpha\in [0,2\pi]$. (For details of the model see Appendix~\ref{app:pol_3b_model}.)
To be concrete, let us consider the lowest two bands $n=1,2$. As can be seen from Fig.~\ref{fig:Vector_field},
$\bar{R}_{12}$ has integer discontinuities at the values of $\alpha$ for which $A_{12}=0$ and $\phi_{12}$ 
is not well defined. The exact location of the discontinuities is determined by the vorticity of the field 
$\v{R}_{12}$ and whether it vanishes or not, see Appendix~\ref{app:pol_3b_model}. 

\section{Two and three dimensions}
We have shown in detail how the integral of the shift vector is related to the 
electric polarization differences in 1D. We next consider generalizations to
higher dimensional. In higher dimensions the shift vector has two or more  
Cartesian indices $a,b=x,y,z$. The analogous definition to Eq.~(\ref{eq:Wcv_def}) is
\begin{align}
W_{nm}^{a,b} =\frac{v}{l^a} \int_{\textit{{\tiny BZ}}} \frac{\partial \phi^b_{nm}}{\partial k^a},
\end{align}
where $l^a$ is the primitive lattice vector component and
$v$ is the volume of the primitive unit cell. 
If we define the integral of shift vector over the BZ as
\begin{align}
\bar{R}^{a,b}_{nm} &= v \int_{\textit{{\tiny BZ}}}  R^{a,b}_{nm},
\end{align}
then we obtain
\begin{align}
e \bar{R}^{a,b}_{nm} = v (P^a_n - P^a_m) +  W^{a,b}_{nm} v Q^a, 
\label{eqn:shift_vec_pol_general}
\end{align}
where $Q^a=e l^a/v$ is the quantum of polarization along the $a$ 
Cartesian axis. There are two situations of interest. First, 
if there are no optical zeros on the parameter space path,
we can define an optical gauge where the polarization difference
can be inferred from the integral of the shift vector with 
$W_{nm}^{a,b}=0$. In this case,  $e \bar{R}^{a,b}_{nm} = v (P^a_n - P^a_m)$ 
holds and materials with large polarization differences (the right hand side) 
lead to efficient photovoltaic responses (through shift vector in the left hand side).

Second, if there exist optical zeros, the optical gauge has discontinuities. 
Then, $W_{nm}^{ab}$ is not quantized since a winding number 
$\int dk_a \partial_{k_a} \phi_{nm}^b$ as a function of $k_c$ ($c \neq a$) in general 
has jumps at optical zeros. In  this case, while we cannot directly relate 
$\bar{R}^{a,b}_{nm}$ and $P^a_n - P^a_m$, the right hand side including $W^{a,b}_{nm}$ 
can be evaluated in a fairly easy way, providing a guideline to search efficient 
photovoltaic materials. In particular, Eq.~(\ref{eqn:shift_vec_pol_general}) shows that 
the polarization difference and locations of optical zeros (that determine $W^{a,b}_{nm}$) 
are important in understanding photovoltaic responses in the left hand side.

As an example, consider a simple extension of the RM model to two dimensions. It consists of two 1D RM 
models, one in the $x$ direction and the other in the $y$ direction, with dimerization parameters, 
$\delta_x,\delta_y$. We suppose that the staggered potential is modulated along $x$ but constant along $y$.
It is easy to show that the electrical polarization is along $x$ and only transitions from bands 
$1\to 3$ and $2\to 4$ are allowed. There is a line of optical zeros at $(k_x,k_y,\delta_x)=(0,k_y,0)$ 
for all $k_y$ and one can define gauge-invariant fields in the plane $(k_x,\delta_x)$ with similar vorticity 
as in Fig.~\ref{fig:shift_vector_RM_model}(c). As long as $\delta_x\neq 0$ the winding $W^{xx}_{12}=0$. 
Similarly to the RM model in 1D, the existence of the singularity at $\delta_x=0$ gives rise to a discontinuity 
of $\bar{R}^{xx}_{12}$.

\section{Discussion and conclusions}
We demonstrated that the integral of the shift vector is a dominant factor in determining the total shift current 
generated in 2D materials. Barring points where the optical transitions are forbidden, the integral of the shift 
vector has the meaning of polarization differences between conduction and valence bands.
We also describe the theoretical tools for analyzing the polarization differences in the presence or absence of 
optical zeros. With the caveats explained above, Eq.~(\ref{supp:shift_tensor_main_text_v3}) gives   

\begin{align}
\int d\omega~ \sigma^{aaa}\approx -\frac{\pi e^2 \varepsilon^{aa}}{\hbar^2}\sum_{nm} f_{nm}(P^a_n - P^a_m),
\label{supp:shift_tensor_main_text_v4}
\end{align}
where we assumed the optical gauge and zero temperature where $f_{nm}=-1$ for $n(m)$ a conduction (valence) 
band and $=1$ when $n(m)$ a valence(conduction) band. The short-circuit current on a device is 
proportional to the sum of polarization differences. Since the electronic part of the spontaneous polarization is 
the sum over all occupied (valence) band polarizations, Eq.~(\ref{supp:shift_tensor_main_text_v4}) suggests 
that 2D ferroelectrics are natural candidates for materials with large shift-current generation. Hence, our 
results provide the long-sought link between electric polarization and shift current.

There is numerical evidence that 2D ferroelectric single-layer IV monochalcogenides have large shift current
~\cite{cook_design_2015,Rangel2017}. A recent experiment measuring shift current on thin films of GeS 
is consistent with our results~\cite{Kushnir2017}. We also expect large shift current in the recently 
discovered 2D ferroelectric SnTe\cite{Chang2016}.  Finally, the right-hand side of Eq.~\ref{eqn:shift_vec_pol_general} 
is easier to evaluate than the left-hand side with standard \textit{ab} \textit{initio} methods and serves as an estimate 
of shift current generation and provides a practical guideline to search for materials with large shift currents. 

\section{acknowledgments}
We thank F. de Juan, M. Kolodrubetz and S. Barraza-Lopez for useful discussions.  B.M.F. acknowledges 
support from AFOSR MURI, Conacyt, and NERSC Contract No. DE-AC02-05CH11231.
T.M. acknowledges support from the Gordon and Betty Moore Foundation's EPiQS Initiative Theory Center Grant.
J.E.M. acknowledges funding from NSF DMR-1507141 and a Simons Investigatorship.

\appendix

\section{Phase-independent expression of shift vector}
\label{app:phase_ind_shiftv}
The shift current tensor, Eq.~(2) in the main text, in $d$ dimension is usually written as~\cite{Sipe2000}
\begin{align}
\sigma^{abc}(0;\omega,-\omega) = \frac{i\pi e^3}{2\hbar^2}&\int_{\textit{{\tiny BZ}}} \sum_{nm} f_{nm}
(r_{mn}^b r_{nm;a}^c \nonumber \\
&+ r_{mn}^c r_{nm;a}^b) \delta(\omega_{mn}-\omega),
\label{supp:shift_tensor_gen}
\end{align}
Here we defined the integral as $\int_{\textit{{\tiny BZ}}}\equiv\int d \v{k}/(2\pi)^d$ over the 
Brillouin zone  (BZ) in d dimensions for notational convenience. 
$\hbar\omega_{nm}=\hbar\omega_{n}-\hbar\omega_{m}$  are band energy differences
and $f_{nm}=f_n - f_m$ differences of Fermi distribution functions of band $n$ and 
$m$. The dipole matrix elements $r_{nm}^a$ and generalized derivatives are
\begin{align}
r_{nm}^b &\equiv A_{nm}^b ~~~ [n\neq m ~~\textrm{and} ~~0~~ \textrm{otherwise}]  \\
r_{nm;a}^b &\equiv \frac{\partial r_{nm}^b}{\partial k^a} -i (A_{nn}^a -A_{mm}^a ) r_{nm}^b.
\end{align}
$A^b_{nm}= i\langle u_n|\frac{\partial}{\partial k^b} |u_m\rangle$ 
are the Berry connections, $\hbar\omega_{nm}= \hbar\omega_n - \hbar\omega_m$ 
are the band energies and $f_{nm}= f_n- f_m$ are the fermionic occupation 
numbers. We can write $A_{nm}^{b} = v_{nm}/i\omega_{nm}$, for nondegenerate bands
where $v^b_{nm}$ is the velocity matrix element. Setting $b=c$ for linear polarization 
and using polar representation, $r^{a}_{nm} = |r^a_{nm}|e^{-i\phi_{nm}^a}$,
Eq.~(\ref{supp:shift_tensor_gen}) reduces to
\begin{align}
\sigma^{abb}(0;\omega,-\omega) = -\frac{\pi e^3}{\hbar^2}\int_{\textit{{\tiny BZ}}} \sum_{nm}& f_{nm}
R_{nm}^{a,b} |r_{nm}^b|^2  \nonumber \\
&\times\delta(\omega_{mn}-\omega),
\label{supp:shift_tensor}
\end{align}
where $R_{nm}^{a,b}$ is the so-called shift `vector,'
\begin{align}
R^{a,b}_{nm} =\frac{\partial \phi^{b}_{nm}}{\partial k^a} + A^a_{nn} -A^a_{mm},
\label{eqn:shift_vec_multiband_supp}
\end{align}
An alternative expression for the shift vector, which avoids the use of $\phi_{nm}^b$, can be obtained from
Eq.~(\ref{supp:shift_tensor_gen}). Since $\sigma_2(0;\omega,-\omega)$ is real we have
\begin{align}
R_{nm}^{a,b}|r_{nm}^b|^2 = -\textrm{Im}\big[r_{mn}^{b}r_{nm;a}^{b}\big].
\label{supp_eq:integrand_shift_current}
\end{align}
The right-hand-side is gauge invariant and has simple analytical expressions 
for effective models of monochalcogenides~\cite{cook_design_2015,Rangel2017}. 
It contains two important physical effects, density of states and the geometry of
Bloch wave functions. To disentangle these effects, let us consider the case where $r_{nm}^b\neq 0$ 
(equivalently $v^b_{nm}\neq 0$) then the shift vector itself is well defined,
\begin{align}
R_{nm}^{a,b} = -\frac{1}{|r_{nm}^b|^2}\textrm{Im}\big[r_{mn}^{b}r_{nm;a}^{b}\big],
\label{eqn:Rnm_def}
\end{align}
and independent of the density of states. In the independent-particle approximation, 
the imaginary part of the dielectric function,
\begin{align}
\frac{\varepsilon^{ab}_2(\omega)}{\varepsilon_0}&= \delta_{ab}-\frac{e^2 \pi}{\varepsilon_0\hbar} \int_{\textit{{\tiny BZ}}} \sum_{nm} f_{nm} r^a_{nm}r^b_{mn} \delta(\omega_{mn} - \omega).
\label{eq:epsilon2}
\end{align}
is dominated by the second term and comparing with Eq.~(\ref{supp:shift_tensor}) we obtain Eq.~(2) in the main text. 

\section{The optical gauge}
\label{app:optical_gauge}
The solutions of the Schrodinger equation with a periodic potential are Bloch wavefunctions,
\begin{align}
\psi_n(\v{k},\v{r}) = e^{i\v{k}\cdot \v{r}}u_n(\v{k},\v{r}), 
\end{align}
where $n$ is the bands index and $\v{k}$ the crystal momentum.
$u_n(\v{k},\v{r}+\v{R})=u_n(\v{k},\v{r})$ is the cell periodic part of the 
wave function and $\v{R}$ is a lattice vector. The solutions of the 
Schrodinger equation are invariant under phase transformations [$U(1)$ gauge transformations],
\begin{align}
\psi'_n(\v{k},\v{r}) = e^{i\varphi_n(\v{k})}\psi_n(\v{k},\v{r}).
\end{align}
Under gauge transformations the Berry connections transform as
\begin{align}
A{'}_{nm}^b &= A_{nm}^b e^{i(\varphi_m-\varphi_n)} \\
A{'}_{mm}^b &= A_{mm}^b - \frac{\partial \varphi_m(\v{k})}{\partial k^b}.
\label{eqn:Anm_phasetrans}
\end{align}
The diagonal matrix elements can change by an arbitrary 
phase $\varphi_n$. Hence choosing the diagonal elements is equivalent to 
fixing a particular gauge. On the other hand, the off diagonal Berry connections 
transform as operators and therefore, if $A_{nm}^b=0$ in one gauge it vanishes in all gauges.
The dipole matrix elements and its generalized derivatives transform as operators 
\begin{align}
r{'}_{nm}^b &= e^{i(\varphi_m-\varphi_n)} ~ r_{nm}^b \\
r{'}_{nm;a}^b &= e^{i(\varphi_m-\varphi_n)} ~ r_{nm;a}^b,
\end{align}
but the standard derivative $\partial r_{nm}^b/\partial k^{a}$ does not transform 
as a tensor. From these results we see that the shift vector, Eq.~(\ref{eqn:Rnm_def}), 
is gauge invariant.

\noindent Now, the Bloch states at $\v{k}$ and $\v{k}+\v{G}$, with $\v{G}$ a reciprocal lattice vector, 
are physically equivalent states. They can differ at most by a phase $\lambda$, 
\begin{align}
\psi_n(\v{k}+\v{G})= \lambda_n \psi_n(\v{k}),
\end{align} 
where $\lambda_n= e^{i\theta_n(\v{k},\v{G})}$ is determined by the choice of $\varphi_n$. 
For arbitrary $\lambda_n$ the connections at $\v{k}$ and $\v{k}+\v{G}$ are related as
\begin{align}
A^b_{mm}(\v{k}+\v{G}) &= A^b_{mm}(\v{k}) + \lambda^{*}_m i \frac{\partial \lambda_m}{\partial k^b}  \\
A^b_{nm}(\v{k}+\v{G}) &= \lambda^{*}_{n} \lambda_m A^b_{nm}(\v{k}).
\end{align}
In general, the off-diagonal elements at $\v{k}$ and $\v{k}+\v{G}$ differ by an arbitrary phase,
but if we choose the periodic gauge where $\lambda_n =1$, then both the Bloch wave functions 
and connections (diagonal \textit{and} off-diagonal) are periodic.
Note that the phases at $\v{k}$ and $\v{k}+\v{G}$ may differ by an 
integer multiple of $2\pi$. The ambiguity in $A_{nn}^b$ gives rise to an integer
ambiguity in the polarization and the ambiguity in $A_{nm}^b$ to 
the interband winding number $W_{nm}^{b,b}$ described in the main text.
This is because we still have freedom to impose gauge transformations 
in which $\nabla_\v{k} \varphi(\v{k})$ is periodic~\cite{Lax1974}, which 
include large gauge transformations. Let us call the subset with $W_{nm}^{a,b}=0$ the optical gauge.

\section{Shift vector of two-band model from Hamiltonian derivatives}
\label{app:two-band_model}
For a two-band Hamiltonian given in first quantization as $H = \sum_i d_i \sigma^i$, where 
$\v{d}=(d_x,d_y,d_z)$, the right-hand side of Eq.~(\ref{supp_eq:integrand_shift_current}) is
\begin{align}
\textrm{Im}\big[r_{12}^{b}r_{21;a}^{b}\big] = \epsilon_{mij} \frac{1}{4E^5} \big(d_m d_{i,a}& d_{j,b} d_l d_{l,a} \nonumber \\
&- E^2 d_{m} d_{i,a} d_{j,ab}   \big).
\end{align}
$\pm E(\v{k})$ are the eigenvalues of the Bloch Hamiltonian, and 
$d_{i,a} = \partial d_i /\partial k^a$. 
This result is easier to obtain by expanding both sides of the identity  
$\partial_{k^b}\partial_{k^a} \langle u_n|H |u_m \rangle = \delta_{nm} \partial_{k^b}\partial_{k^a}  E_{n}$.
From this we obtain an expression for the generalized derivative in terms of 	
velocity matrix elements only~\cite{cook_design_2015,Sipe2000}, ($ n \neq m$)
\begin{align}
r^a_{nm;b}  = -\frac{1}{i\omega_{nm}} \bigg[ &\frac{v^a_{nm}\Delta^b_{nm} +v^b_{nm}\Delta^a_{nm}}{\omega_{nm}} 
-w_{nm}^{ab} \nonumber \\
&+ \sum_{p\neq n, m}(\frac{v^a_{np} v^b_{pm}}{\omega_{pm}}- \frac{v_{np}^b v^a_{pm}}{\omega_{np}})\bigg], 
\end{align}
where $v^b_{nm} = \left<n|\partial_{k_b}H|m\right>$ are the velocity matrix 
elements, $\Delta^b_{nm} = v^b_{nn} - v^b_{mm}$, $w_{nm}^{ba}=\left<n|\partial_{k_b} \partial_{k_a}H|m\right>$ 
and $\hbar\omega_{nm} = E_n - E_m$. In the evaluation, we used various standard identities.
Note the extra term $w_{nm}^{ab}$ compared to Ref.~\cite{Sipe2000}, 
where $H = p^2/2m +V(x)$ and $w_{nm}^{ab}=\delta_{nm}\delta^{ab}/m$ is diagonal.
Tight-binding models are, of course, approximations to real-life solid 
state Hamiltonians and comparison with experiments must proceed with caution to 
avoid spurious terms arising from the use of a tight-binding model.
$r^b_{nm}$ can also be obtained in terms of Hamiltonian derivatives.
Recall that by definition only off-diagonal terms contribute,
\begin{align}
|r_{12}^b|^2 = \frac{1}{4 E^2 (E^2 - d_z^2)}&\big[ (d_z E_{,b} -d_{z,b} E)^2 \nonumber \\ 
&+ (d_x d_{y,b} - d_{x,b} d_{y} )^2  \big].
\end{align}
Hence the shift vector written as 
\begin{align}
R_{12}^{a,b} = - \epsilon_{mij}  \frac{(E^2 \hspace{-2pt}-\hspace{-3pt} d_z^2) (d_m d_{i,a} d_{j,b} d_l d_{l,a} 
\hspace{-3pt}-\hspace{-3pt} E^2 d_{m} d_{i,a} d_{j,ab})}{E^3 \big[ (d_z E_{,b} -d_{z,b} E)^2 \hspace{-3pt}+ \hspace{-3pt}(d_x d_{y,b} - d_{x,b} d_{y} )^2  \big] } 
\label{eqn:R12_2B_model}
\end{align}
is explicitly gauge independent. In particular, the expression for the shift 
vector for $b=a$ reduces to
\begin{align}
R_{cv}^{a,a} = 
-\frac{| \v{d}|  \v{d} \cdot ({\v{d}'} \times {\v{d}''})} {| \v{d}|^2 |\v{d}'|^2 - (\partial_{k^a} |\v{d}|^2)^2/4},
\label{eqn:Rcv_RM_compact}
\end{align}
where $d'_i= \partial_{k^a} d_i$. 

\begin{figure*}[]
\subfigure{\includegraphics[width=.9\textwidth]{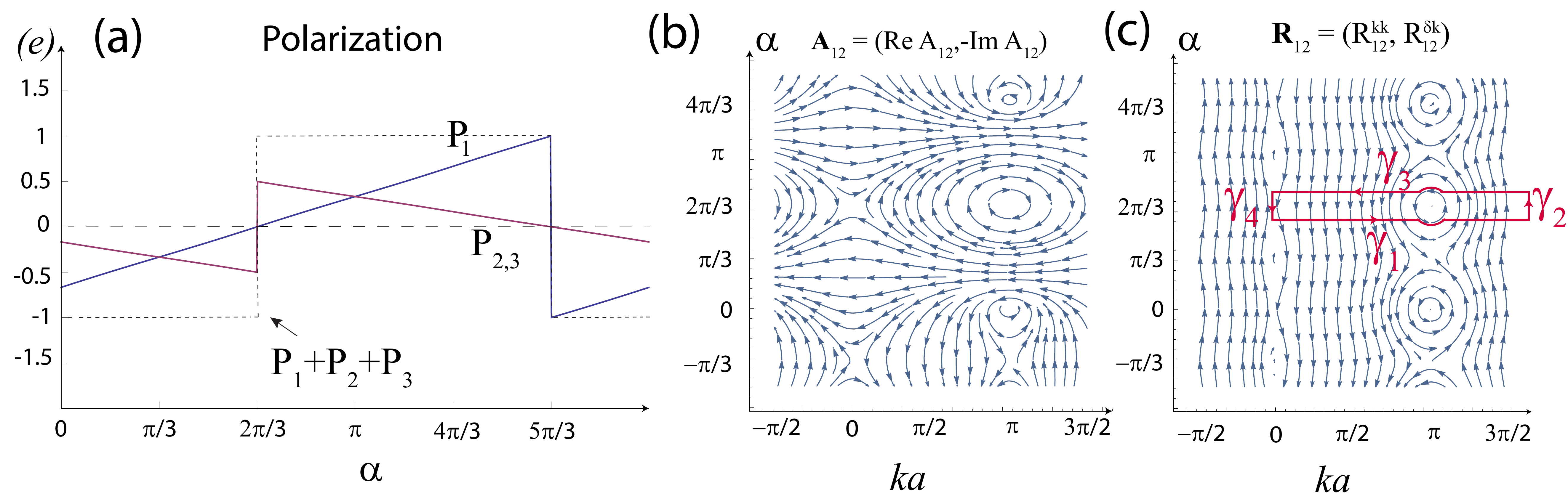}}
\caption{Color online. (a) polarizations of each band of model Eq.~(\ref{sup:eq:3B_model_ham})
as a function of $\alpha$. (b) $W_{12}$ changes at optical zeros $\alpha=0,4\pi/3$ and at 
the inversion symmetric point $\alpha=5\pi/3$. (c) Gauge-invariant field $(R^{k k}_{12}, R^{\delta k}_{12})$ 
showing the vorticity of the optical zeros giving the discontinuities of $\bar{R}_{12}$. 
The loop $\gamma = \sum_n \gamma_n$ encloses a vortex of charge +1 (see main text). 
One can check that $\v{R}_{12}=0$ at $ka=0$, and hence it does not contribute to the path 
integral. We chose units such that $e=a=1$.}
\label{fig:Vector_field_3B_model}
\end{figure*}

\section{Shift vector and current in Rice-Mele model}
\label{sec:shift_vector_RM}
In this section the shift vector and shift current for the 
Rice-Mele model of ferroelectrics is computed. The Hamiltonian is

\begin{align}
\hat{H}_{RM}= \sum_{i} [ (\frac{t}{2} + (-1)^{i} \frac{\delta}{2}) (c_{i}^{\dagger} c_{i+1} + h.c.) + (-1)^{i} \Delta c^{\dagger}_{i} c_{i} ],
\end{align}
where $c_i$($c^{\dagger}_i$) destroys (creates) electron states at site $i$, $\delta$ parametrizes 
the dimerization of the chain, and $\Delta$ is the staggered potential on sites A and B. If $\Delta\neq 0$, 
and $\delta \neq 0$ inversion symmetry is broken. The unit cell (of length $a$) has two sites.  
We obtain the Bloch  Hamiltonian, 
\begin{align}
H_{RM}= \sum_i d_i \sigma_i = \sigma_x~ t\cos k a/2  - \sigma_y~ \delta \sin ka/2  + \sigma_z ~\Delta
\end{align}
and eigenfunctions,

\begin{align}
u_c=\frac{e^{i\varphi_{c}}}{\sqrt{2}}  
\begin{pmatrix}
    v      \\
    u e^{ i \phi}
\end{pmatrix}
\hspace{15pt}
u_v=\frac{e^{i\varphi_{v}}}{\sqrt{2}} 
\begin{pmatrix}
    u      \\
    -v e^{ i \phi}
\end{pmatrix},
\label{eq:wf_RM_model}
\end{align}
where $H_{RM}u_{c,v} = \pm E u_{c,v}, \phi = \arctan[(-\delta/t) \tan(ka/2)]$ (mod $\pi$) 
is the azimuthal angle of the vector $H_{RM}$ in the Bloch sphere, $u=\sqrt{1-\Delta/E}$, 
$v=\sqrt{1+\Delta/E}$, and the eigenvalues are given by $E=(t^2 \cos^2 ka/2 + \delta^2 \sin^2 ka/2 + \Delta^2)^{1/2}$ 
for the conduction and $-E$ for the valence band ($\phi$ should not be confused with $\phi_{cv}$). 
We have added a gauge dependence $\varphi_{n}$, ($n=c,v$). The Berry connection will depend 
explicitly on the gauge used but results on the shift vector/current are gauge 
independent. In this section we choose $\partial_k \varphi_{n}=0$. The Bloch wave functions are 
$\psi_{n}(k,r)= \sum_{j} e^{i k a j} [u^{A}_{n}(k)\chi(r-a j) + e^{i k a /2} u^{B}_{n}(k)\chi(r-a j-a/2)]$,
where $\chi$ are the atomic wave functions and $u_{n}^{A,B}$ projections of the eigenfunctions on site A(B). 
The Berry connections,
\begin{align}
A_{nn} &= i u_{n}^{\dagger} \partial_k u_{n} = \frac{a t\delta (E \mp \Delta) }{4 E(E^2- \Delta^2)} ~~~~ (n =c,v)\\
A_{cv} &= i u_c^{\dagger} \partial_k u_{v} =\frac{a ~i e^{- i(\varphi_{v}-\varphi_{c})} }{ 8 E^2 \sqrt{E^2- \Delta^2}} \big[\Delta (t^2-\delta^2)\sin ka \nonumber \\
&\hspace{155pt}+ 2 i \delta t E\big],
\label{eq:diag_berry_con}
\end{align}
are both periodic with period $2\pi/a$. We define the phase $\phi_{cv}$ 
by $A_{cv }= |A_{ cv}| e^{-i \phi_{cv}} = |A_{cv}|(\cos \phi_{cv},-\sin\phi_{cv})$,
and its derivative is,	
\begin{align}
\partial_k \phi_{cv} =& 
\frac{\Delta}{ 2E} \frac{a \delta t  (\delta^2 \hspace{-3pt}-\hspace{-3pt} t^2) \left[4 E^2 \cos k a \hspace{-3pt}+\hspace{-3pt} (t^2-\delta^2)\sin^2 ka\right]}{[\Delta^2 (\delta^2 - t^2)^2\sin^2 ka + 4 \delta^2 t^2 E^2]}.
\end{align}
This expression is smooth for $\delta \neq 0$. If $\delta=0$ it can be 
seen that $A_{cv}=0$ at $ka=0,\pi$. The shift vector, $R_{cv}=  \partial_k\phi_{cv} + A_{cc} -A_{vv}$,
can be computed analytically as
\begin{align}
R_{cv}= &\frac{\Delta}{2E} \frac{a \delta t  (\delta^2 - t^2) \big[4 E^2 \cos k a + (t^2-\delta^2)\sin^2 ka\big]}{[\Delta^2 (\delta^2 - t^2)^2\sin^2 ka + 4 \delta^2 t^2 E^2]} \nonumber\\
&\hspace{110pt} - \frac{\Delta}{2 E} \frac{a t\delta }{(E^2- \Delta^2)}.
\end{align}
Some observations about the behavior of the shift vector in the RM model are in order: (a) Generally, 
the shift vector does not vanish. (b) The shift vector is peaked at $ka =0$ or $ka=\pi$,
and (c) the shift vector can exceed the lattice spacing $a$.
To illustrate this consider some limiting values of the shift vector,
\begin{align}
&R_{cv}\big|_{\textrm{lim} ~k\to 0 } = -\frac{a t \Delta}{2\delta \sqrt{t^2 + \Delta^2}}\big|_{\textrm{lim}~\delta \to 0}= \infty 
\label{eq:Rcv_infinity}\\	
&R_{cv}\big|_{\textrm{lim} ~k\to \pi/a} = -\frac{a \delta \Delta}{2t \sqrt{\delta^2 + \Delta^2}}\big|_{\textrm{lim}~\delta \to 0}= 0  \\
&R_{cv}\big|_{\textrm{lim} ~\delta \to t} = -\frac{a \Delta}{2 \sqrt{t^2 + \Delta^2}}\big|_{\textrm{lim}~\Delta \to 0} = -a\Delta/2t 
\nonumber\\
 &\hspace{110pt} (\textrm{flat band limit}).
\end{align}
Hence one can check that at $ka=\pi$ the field $\v{R}_{cv} = (R_{cv}^{kk},R_{cv}^{\delta k})$, defined in the main text,
vanishes and at $ka=0$ it diverges. 

\noindent \textit{Shift current}.-
If the electric field is along the chain, e.g., the $z$ direction, 
the shift current is  
\begin{align}
J^{z}_{\textrm{shift}}(\omega) =  2\sigma^{zzz}(0;\omega,-\omega) E^z(\omega)E^z(-\omega).
\end{align}
For the two-band model this reduces to 
\begin{align}
\sigma^{zzz}(0;\omega,-\omega) =  e^3 \int_0^{2\pi/a} d k \frac{|\langle u_c |v^z| u_v \rangle|^2 R_{cv} }{\hbar^2\omega^2}  \delta(\frac{2E}{\hbar} -\omega)
\end{align}
where the matrix elements of the velocity operator $v^z= \hbar^{-1}\partial H_{RM}/\partial k $ are 
\begin{align}
|\langle u_c |v^z | u_v \rangle|^2 = \frac{a^2}{16 \hbar^2} \frac{1}{E^2 (E^2 \hspace{-3pt}-\hspace{-3pt} \Delta^2)} &\big( \Delta^2 (t^2\hspace{-3pt}-\hspace{-3pt}\delta^2)^2 \sin^2 ka \nonumber \\
&\hspace{20pt}+ 4 t^2 \delta^2 E^2 \big).
\end{align}
The shift vector and the matrix elements of the velocity each have complicated expressions but the 
combination (the `integrand'), 
\begin{align}
\frac{\hbar^2 |v^z_{cv}|^2}{4 E^2} R_{cv} = |r^z_{cv}|^2 R_{cv} = -\textrm{Im}[r^z_{cv}r^z_{vc;z}],
\end{align}
is simply 
\begin{align}
\textrm{Im}[r^z_{cv}r^z_{vc;z}] = \frac{a^3 t \delta \Delta}{32 E^3}.
\end{align}
For $\delta \ll \Delta$, $R_{cv}$ is sharply peaked at $ka=0$ but $|r^z_{cv}|$ peaks at $ka=\pi$.
As $\delta$ increases the peak in $R_{cv}$ and $|r^z_{cv}|^2$ broadens but their peaks' maximum also decreases.
The dependence on the velocity matrix elements (imaginary part of the dielectric function) is very prominent 
here because the system is 1D. The analytical expression for the shift current of the 
RM model simplifies to 
\begin{align}
\sigma^{zzz}(0;\omega,-\omega) =& -\frac{ e^3 a^3 t \delta  \Delta }{8	 \hbar^4 \omega^3 } \sum_{i} \frac{1}{|\partial_k E(k_i)|	},
\end{align}
where $\partial_k E= a(\delta^2-t^2) \sin k a/4E$ is the velocity at 
momentum $k$ and $k_i$ are the two solutions of $2E(k_i)=\hbar\omega$ for $\hbar \omega > 2E$.
In 1D, $\sigma^{zzz}$ diverges as $\omega^{-3} (2E -\hbar \omega)^{-1/2}$ at the band edge, but 
is suppressed in 2D, where the role of the shift vector becomes prominent.

\section{Polarization and shift vector in a three-band model}
\label{app:pol_3b_model}
Let us consider the 1D model Hamiltonian

\begin{align}
\hat{H}_{\textit{{\tiny 3B}}}=\sum_j t_j c^{\dagger}_j c_{j+1} + h.c.,
\label{sup:eq:3B_model_ham}
\end{align}
with $t_j= A+ B \cos{\big(2\pi j/3 -\alpha\big)}$. There are three distinct values of the hoppings $t_j=t_1,t_2,t_3$. 
Hence, the unit cell (of size $a$) has three nonequivalent sites. The crystal has inversion symmetry for 
$\alpha=0,\pi/3, 2\pi/3, \pi, 4\pi/3, 5\pi/3$, when two of the hoppings are equal. The Bloch Hamiltonian is
\begin{align}
H_{\textit{{\tiny 3B}}}=
\begin{bmatrix}
    0       & t_1 e^{i ka/3} &  t_3 e^{-i ka/3} \\
    t_1 e^{-i ka/3} & 0 &  t_2 e^{i ka/3} \\
    t_3 e^{i ka/3}  & t_2 e^{-i ka/3} & 0
\end{bmatrix}.
\end{align}
The eigenvalues and eigenvectors, $H_{\textit{{\tiny 3B}}} u_n = E_n u_n$, are  
\begin{align}
E_{n} = 2 t_{r} \cos{\big(\frac{1}{3}\arccos{(t_g^3 \cos{(k a)}/t_r^3)}- \frac{2\pi n}{3} \big)},
\label{eq:banddisp}
\end{align}
where $(n=1,2,3)$ and we defined the root mean square and geometric average  
$t_r = \sqrt{(t_1^2 +t_2^2 +t_3^2)/3}$ and $t_g = (t_1 t_2 t_3)^{1/3}$ respectively, and 
\begin{align}
u_n=\frac{e^{i\varphi_{n}}}{N_n}
\begin{bmatrix}
    E_n^2 - t_2^2   \\
    t_2 t_3 e^{2 i ka/3} + E_n t_1 e^{- i ka/3}   \\
    t_2 t_1 e^{-2 i ka/3} + E_n t_3 e^{i ka/3}
\end{bmatrix},
\label{eq.wave_func}
\end{align}
where $\varphi_{n}$ is chosen to enforce the periodic gauge. The normalization is 
$N_n= [(E_n^2 - t_2^2)^2 + (t_2^2 - E_n^2)(t_3^2 - E_n^2) +(t_1^2 - E_n^2)(t_2^2 - E_n^2)]^{1/2}$.
Using these wave functions, the Berry connections are calculated analytically as,
\begin{align}
A_{nn} & = \frac{a}{2 N_n^2} (E_n^2 + 2t_2^2)(t_1^2 - t_3^2) \nonumber \\
&- \frac{i}{N_n^2}\big[ 3 E_n (\partial_k E_n) (E_n^2 - t_r^2) + 2aE_n t_g^3 \sin{(k a)}\big]\\
A_{nm} & = \frac{a}{3 N_n N_m} (E_n E_m +2 t_2^2) (t_1^2-t_3^2) \nonumber \\
& + \frac{i 2 a t_g^3 (E_n^2 + E_m^2 + E_n E_m -3 t_2^2) \sin{(k a)} }{3 N_n N_m (E_n-E_m)}.  
\label{eq:berry_con1}
\end{align}
Note that the Berry connections are periodic in $k$ space with period $G=2\pi/a$.
One can check that the optical zeros of $A_{12}$ are $\alpha=0, 2\pi/3, 4\pi/3$, 
where $A_{12}(ka=\pi,\alpha)$ vanishes and  hence the phase of 
$\phi_{12}$ is not well defined. 
In Figs.~\ref{fig:Vector_field_3B_model}(a) and (b) we show the windings of the interband
connection $A_{12}$ and the gauge-invariant vorticity of the optical zeros in 
the field $\v{R}_{12}$ described in the main text.

\noindent \textit{Polarization}.-
The polarization is given by the integral over the Berry connection as
\begin{align}
P_n(\alpha) = \frac{1}{2\pi} \int dk A_{nn}(k,\alpha).
\label{eq:polarization_def}
\end{align}
In Fig.~\ref{fig:Vector_field_3B_model}(a) we show the individual band polarizations as a function of 
$\alpha$. Note that the sum $P_1 + P_2 + P_3 = \pm 1$ ~(mod $e$), as expected.
Also, the total charge pumped of band $n$ per cycle is 
\begin{align}
c_n(\alpha) &= \int_0^{\alpha} d\lambda \int \frac{dk}{2\pi} \Omega^{n}_{k,\lambda} \nonumber \\
&= \int_0^{\alpha} d\lambda \int \frac{dk}{2\pi} i \big[ \langle \partial_k u_n | \partial_\lambda u_n\rangle - \langle \partial_\lambda u_n | \partial_k u_n\rangle\big]
\label{eq:charge_pumped}
\end{align}
One can check the charge pumped across the unit cell is $c_1(2\pi)=-2e$ and $c_{2,3}(2\pi)=+e$.

%

\end{document}